\documentclass{revtex4}

\usepackage[latin1]{inputenc}
\usepackage[english]{babel}
\usepackage{graphicx}
\usepackage{amsfonts,amsmath,amssymb}
\usepackage{hyperref}
\usepackage{feynmf}
\usepackage{color}

\newenvironment{arbre}{\begin{minipage}[c]{3.2cm}
\bigskip \begin{center} \begin{fmfgraph*}(50,50)}{ \end{fmfgraph*}
\end{center} \bigskip \end{minipage}}

\begin{document}

\title{Evolution of the most recent common ancestor of a population with no selection}

\author{Damien Simon}
\email{damien.simon@lps.ens.fr} \affiliation{Laboratoire de
Physique Statistique,\\ \'Ecole Normale Sup\'erieure,\\24, rue
Lhomond, 75231 Paris Cedex 05, France}

\author{Bernard Derrida}\email{bernard.derrida@lps.ens.fr}
\affiliation{Laboratoire de Physique Statistique,\\ \'Ecole
Normale Sup\'erieure,\\24, rue Lhomond, 75231 Paris Cedex 05,
France}

\date{\today}

\begin{abstract}

We consider the evolution of a population of fixed size with no selection. The number of generations $G$ to reach the first common ancestor evolves in time. This evolution can be described by a simple Markov process which allows one to calculate several characteristics of the time dependence of $G$. We also study how $G$ is correlated to the genetic diversity.

\end{abstract}

\maketitle 

                                         \section{Introduction}
\label{subsec:intro}

One of the simplest questions one can ask about the history of an evolving population is the age of its most recent common ancestor (MRCA). As the population evolves, the age of this MRCA as well as the genealogical tree keep changing with an endless appearance of new branches and disappearance of old branches. These perpetual changes in the genealogy are accompanied by sudden jumps of the age of the MRCA, which correspond to the extinction of one of the oldest branches \cite{serva}. In the first part of the present paper we try to describe the evolution of this age in one of the simplest models of an evolving population, the Wright-Fisher model \cite{wright,fisher, hein, durettbook} with no selection.

Analysis of the human genome makes possible the precise comparison of the DNA sequences of individuals in a population. The number of differences between the sequences of a group of individuals is a testimony of the time passed since their common ancestors and one may hope to infer the history of the group from the knowledge of its DNA sequences \cite{tavareDNA,slatkin,fuli}. The task is however immense as many factors interfere~: selection \cite{kaplan}, history, demography \cite{slatkin,fuli}, geography \cite{wilkinson,wakeleyaliacar}, diploidy \cite{chang,manrubia}. In order to attack the problem of estimating the age of the MRCA from the observed DNA sequences at a given generation, a number of models have been studied \cite{tavareDNA,fuli}, where at most few of these factors are included. The goal is always to correlate the observed genetic diversity of the population at a given generation to the age of this MRCA. However it is difficult to characterize a sample of DNA sequences by a single parameter which would measure its genetic diversity. Ideally the optimal parameter would be to find a measure of the genetic diversity at a given generation which would be as correlated as possible to the age of the MRCA. In practice, one often uses Tajima's estimator \cite{tajima} which counts the number of different base pairs between pairs of individuals. But the more precise the characterization of the genetic diversity is, the more difficult the calculations are \cite{tavareDNA}. Here we consider in the second part of this paper the simple case of the infinite allele model, where the only information we keep about pairs of individuals is whether they have the same allele or not and we try to calculate how the distribution of the age of the MRCA is correlated to this information.

The simplest models one can consider consist in defining some stochastic rules which relate each individual (and its genome) to its parent in the previous generation and the above questions can be formulated as steady state properties of simple non equilibrium systems~: for example the coalescence process described below can be viewed as a reaction-diffusion process $A+A\rightarrow A$. The coalescing trees observed in genealogies have also striking similarities with the ultrametric structures which emerge in the theory of spin glasses and disordered systems \cite{derridapeliti,derrida88}. This is why they motivate a growing interest among statistical physicists \cite{sella}.

We consider here a population of $N$ individuals evolving according to the
Wright-Fisher model (see \cite{hein} for a general introduction)~: successive generations do not overlap, at each new generation all the individuals are replaced by $N$ new
ones and each individual has one parent chosen randomly in the
previous generation. 
	
Many results are known in absence of selection, such as the distribution of the age of the MRCA \cite{kingman, kingman2}, the stochastic dynamics of the frequency of a gene \cite{kimuraa, kimurab}. In the last part of this introduction, we recall few known results that we will use in the rest of the paper. 

Recently Serva adressed the problem of the temporal dynamics of the age of the MRCA. In section \ref{jumps}, we show how to describe these dynamics as a simple Markov process which allows one to calculate all the correlations between these MRCA ages at different generations.

One can associate to each individual a gene (or a genome). In the section \ref{corrdiv} where we try to correlate the genetic diversity to the age of the MRCA, we will consider the infinite allele case~: each mutation creates a new genome, different from all the genomes which had previously appeared in the whole history of the population. At each generation, there is a probability $\theta/N$ of mutation in the transmission of each genome. This means that each new individual inherits the genome of its parent with probability $1-\theta/N$ and receives a new genome with probability $\theta/N$. On average, there are of course $\theta$ mutants in the whole population at each generation. The assumptions made in the infinite allele model and their
links to phylogenetics are discussed in \cite{durettbook} and
\cite{hein}~: it is an approximation which neglects in particular the possibility that two mutations occur on the same base pair.

The results presented in this article are mostly derived in the limit of a large population. It is well known \cite{hein} that, for large $N$, all the relevant times in the genealogy (like for example the age of the MRCA) scale like $N$. In the rest of this paper, we will therefore count the number $G$ of generations in units of $N$ and define the time by $t=G/N$. 

In the remaining part of this introduction we recall some well known properties of the Wright-Fisher model that we will use later \cite{hein}. If one considers a finite number $n$ of individuals, the probability
that these individuals have only $p$ parents in the previous
generation and that they undergo $m$ mutations scales as
$1/N^{(n-p)+m}$~: therefore if one goes back one generation, there is a probability $1-(n(n-1)/2+n\theta)/N$ that the $n$ individuals have different parents and that their genomes are identical to those of their parents. Moreover, there is a probability $n\theta/N$ of observing a single mutation among these $n$ individuals and there is a probability $n(n-1)/(2N)$ that two among the $n$ individuals have the same parent. Therefore, when the size $N$ of the population is large and for $n\ll N^{1/2}$, only pairs of branches coalesce along the tree. The time $T_n$ to find  the Most Recent Common Ancestor to these $n$ individuals can be written as a sum of $n$ independent times $\tau_i$~:
$$ T_n = \tau_2+\tau_3+\ldots +\tau_n$$
where $\tau_i$ is the time spent between the $i^\text{th}$ and the $(i-1)^\text{th}$ coalescence on the tree. This allows one to calculate the distributions $\rho_i(\tau_i)$, as shown in appendix \ref{measgenstruct}~:
\begin{equation}\label{def:rhoi}
\rho_i(\tau_i) = c_i e^{-c_i\tau_i}
\end{equation}
where the coefficients $c_i$ are defined by~:
\begin{equation}\label{def:ci}
c_i= \frac{i(i-1)}{2}
\end{equation}

\begin{figure}
\begin{center}
\input{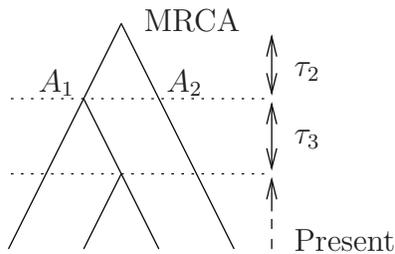}
\end{center}
\caption{\label{fig:arbre}Top of the genealogical tree of a large population. When the size $N$ of the population is large, coalescences at the top of the tree occur only among pairs of individuals and the coalescence times $\tau_i$ are independent random variables. One can see that the two ancestors $A_1$ and $A_2$ generate all the population in the present generation.}
\end{figure}

The generating function of the coalescence time $T_n$ is therefore~:
\begin{equation}\label{qknoinfo} \langle
e^{-\lambda T_n}\rangle =\prod_{i=2}^n\frac{c_i}{\lambda+c_i}\end{equation}
From (\ref{qknoinfo}), one can get the average and the variance of $T_n$~:
\begin{equation}\label{tmoy_unknown}
\langle T_n \rangle =2\Big(1-\frac{1}{n}\Big)\quad\text{and}\quad \langle T_n^2\rangle -\langle T_n\rangle^2= \frac{8}{n}-\frac{4}{n^2}-12 +\sum_{j=1}^n\frac{8}{j^2}
\end{equation}
One can notice that the distribution of $T_n$ remains broad even for large $n$. Although the expressions (\ref{tmoy_unknown}) are derived for fixed $n\ll N$ and in the limit $N\rightarrow\infty$, the
limit $n\rightarrow\infty$ in (\ref{qknoinfo}) and (\ref{tmoy_unknown}) coincides with what would be obtained
by setting $n=N$, i.e. by considering the time $T$ to find the MRCA of the whole population~:
\begin{equation}\label{qpopnoinfo}
\langle e^{-\lambda T}\rangle = \prod_{l=2}^\infty \frac{c_l}{\lambda+c_l}
\end{equation}
It leads to the following expressions of the first two moments of
the coalescence time~:$$ \langle T\rangle = 2
\quad\text{and}\quad \langle T^2\rangle -\langle T\rangle ^2 =
\frac{4\pi^2}{3}-12\simeq 1.159\ldots $$ and to the following stationary distribution
$\rho_\text{st}(T)$~:
\begin{equation}\label{distristatT}
\rho_\text{st}(T)=\sum_{p=2}^\infty (-1)^p(2p-1)c_p e^{-c_p
T}\end{equation}

On the other hand, the stationary distribution of the genomes is
given in \cite{durettbook}~: the probability that, among $n$
individuals, the first $n_1$ have the same genome, the next $n_2$
another genome, and so on until the last $n_k$ which have the
$k^\text{th}$ genome, is given by Ewens' sampling formula \cite{ewens}~:
\begin{equation}\label{Pgroups} P_\text{groups}(n_1,\ldots,n_k) =
\frac{\Gamma(2\theta)}{\Gamma(n+2\theta)}(2\theta)^k\frac{n!}{n_1n_2\ldots
n_k}
\end{equation} where $\Gamma(x)$ is the Euler $\Gamma$ function and $\theta$ the mutation rate.

There are several approaches to calculate the statistical properties of the above model~: either one can write recursive
equations between successive generations and try to solve them, or one can count directly all the possible coalescences and mutations histories of a group. The first approach leads to a
hierarchy of equations, whereas the second option reduces to a simple enumeration. Depending on which of these two approaches appeared to us the simpler to implement, we use alternatively both of them in the present paper. A
coalescence history as described in appendix \ref{measgenstruct} consists in a tree structure, in which each
step corresponds to a coalescence of two individuals chosen
randomly among the $n'\leq n$ which remain, and in a set of $n-1$
times $\tau_i$ between two successive coalescences. A very important simplification (shown in appendix \ref{measgenstruct}) which we will use over and over is that the shape (i.e. the topology) of the trees and the times $\tau_i$ are independent random variables.

%-------------------------------------------------------------------------------------------------------

                 \section{Statistics of the discontinuities of the coalescence time of the population}\label{jumps}

\subsection{Numerical Simulations}

\begin{figure}
\begin{center}
\input{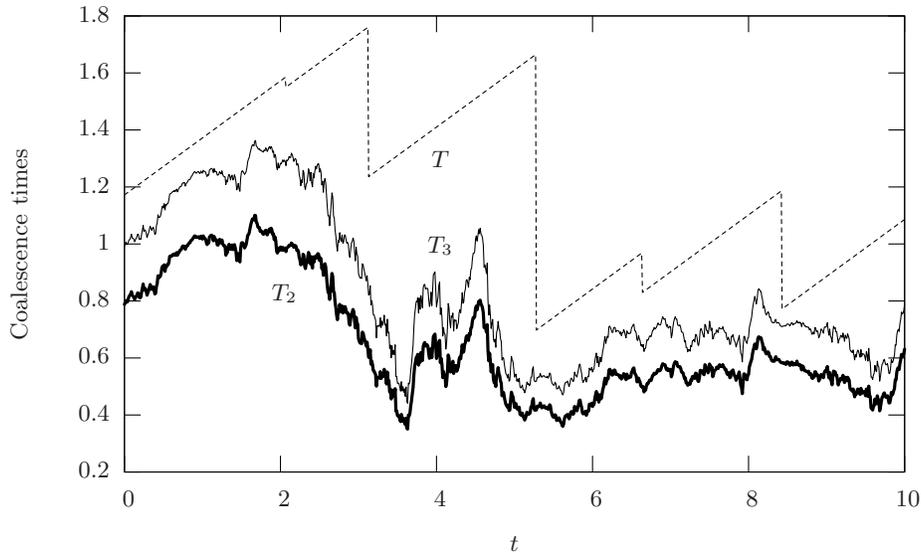}
\end{center}
\caption{Evolution of the age $T=G/N$ of the MRCA for a population
of $N=500$ individuals in the Wright-Fisher model over a rescaled duration $\Delta t=10$, i.e. over $N\Delta t=10N$ generations (dashed line). Thick line~: average $T_2$ over the whole population of the coalescence time of two individuals~; thin
line~: average $T_3$ over the whole population of the coalescence time of
three individuals. One can see that discontinuities are
anticipated by the decreases of the average coalescence time of two
or three individuals.} \label{fig:jumpfig}
\end{figure}

\begin{figure}
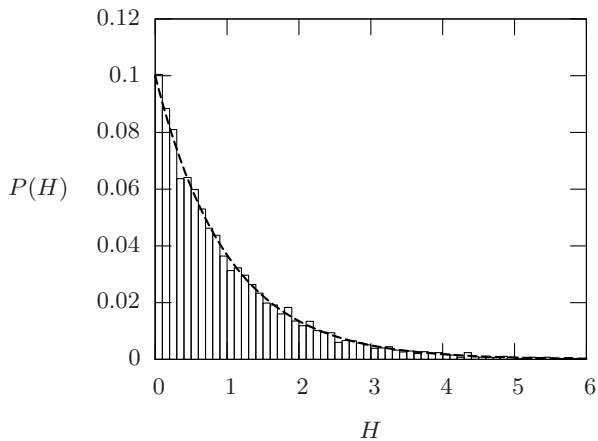

\begin{center}
\input{delays.tex}

\input{heights.tex}
\end{center}
\caption{Measured distributions of the discontinuities of the age
$T$ of the MRCA for a sample of 9169 discontinuities when the
population size is $N=500$ individuals. Top~: histogram of the
distribution of the delays $D_k$ between two successive discontinuities~; the
dashed line is the exponential distribution (\ref{D:exp}).
Bottom~: histogram of the distribution of the jumps $H_k$ at the
discontinuities of $T$~; the dashed line is the exponential
distribution (\ref{H:exp}).}\label{fig:jumpstatfig}
\end{figure}

The Wright-Fisher model implemented for a population of $N=500$
individuals shows interesting features for the evolution of the
coalescence time $T$ (see figure \ref{fig:jumpfig} for $G=5000$ generations,
corresponding to a normalized duration of $\Delta t=10$). The evolution
shows periods of linear increase, separated by discontinuous drops. Let us call $D_k$ the duration of the
$k^\text{th}$ linear increase and $H_k$ the height of the
drop following it. The distributions of the $D_k$'s and $H_k$'s, measured over 9169 discontinuities, 
are shown in figure \ref{fig:jumpstatfig}. Similar results were previously reported in \cite{serva}.

The data of figure \ref{fig:jumpstatfig} indicate that the delays $D_k$ and the heights $H_k$ have an exponential distribution of average 1. The correlations can also be measured (error
bars of order of 0.01)~:
\begin{subequations}
\begin{eqnarray}
\langle D_k D_{k+1} \rangle - \langle D_k\rangle\langle D_{k+1}\rangle &\simeq& -0.005 \label{Dautocorr}\\
\langle H_k H_{k+1} \rangle - \langle H_k\rangle\langle H_{k+1}\rangle &\simeq& -0.006 \label{Hautocorr}\\
\langle H_{k-1}D_k \rangle - \langle H_{k-1}\rangle\langle D_k\rangle &\simeq& -0.002 \label{HDcorr}\\
\langle  H_k D_k     \rangle - \langle H_k\rangle\langle
D_k\rangle
&\simeq& 0.84 \label{DHcorr} \\
\langle H_k D_{k-1}     \rangle - \langle H_k\rangle\langle
D_{k-1}\rangle &\simeq& 0.12 \label{DHcorr2}
\end{eqnarray}
\end{subequations}
This indicates that the only correlation seems to be between the $H_k$ and the previous $D_k$. We try to understand these correlations below.

\subsection{Distribution of delays between two discontinuities}
\label{subsec:jumpstheo}

When $N$ is large, simultaneous coalescence between groups of three or more individuals are negligible (order $1/N^2$) at the top of the tree (i.e. for the last $n$ coalescences with $n\ll \sqrt{N}$, only coalescences of pairs occur). Thus, as shown in figure \ref{fig:arbre}, all the population in the present generation is generated by the two individuals $A_1$ and $A_2$ reached at the penultimate coalescence and thus it can be divided into two groups according to these two ancestors. A discontinuity appears in the age of the MRCA when one of the
two groups generated by $A_1$ and $A_2$ has no offspring. The dynamics of the sizes $N_i$ was studied by Serva in \cite{serva} who showed numerically that the delays $D_i$ have an exponential distribution~:
\begin{equation}\label{D:exp}p_\text{delay}(D)=e^{-D}\end{equation} consistent with the
results of figure \ref{fig:jumpstatfig}. 

\begin{figure}
\begin{center}
\input{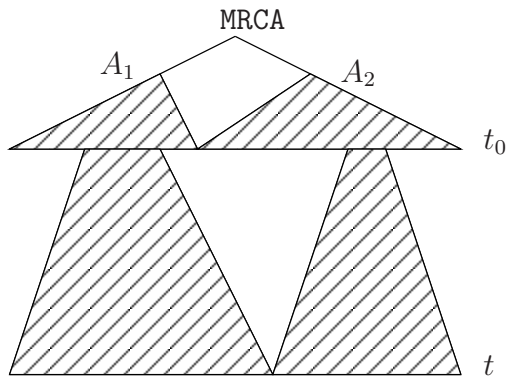}
\end{center}
\caption{\label{fig:arbrett0}Structure of the genealogical tree of the population when the MRCA is the same at $t_0$ and $t$. The population at $t$ must have ancestors at $t_0$ in each of the two groups generated by $A_1$ and $A_2$.}
\end{figure}

%------------------------------------------------------------------------
In order to derive (\ref{D:exp}), let us introduce the probability $P_\text{same}(t_0,t)$ that the MRCA of a population is the same at time $t_0=0$ and at time $t$ (with $t>t_0$), as in figure \ref{fig:arbrett0}. 

As explained above, the population at time $t_0$ can be divided into two parts of size $N_1=x N$ and $N_2=(1-x)N$ according to the ancestors $A_1$ and $A_2$ from which they come. 
The sizes of these two groups are $N_1$ and $N_2=N-N_1$ and one can define the densities
$x=N_1/N$ and $1-x=N_2/N$. At a given generation, $x$
is a random variable in $[0,1]$. Its stochastic evolution is given
by Wright-Fisher rule (see \cite{serva} for an analogy with
brownian motion and its stationary distribution $\rho (x)$ is
uniform on $[0,1]$ for $x$ of order 1 (see \cite{serva} or
appendix \ref{measgenstruct} for a short derivation). There
are finite size correction to this uniform distribution near the boundaries for $x=O(1/N)$ and $1-x=O(1/N)$~; we will not discuss them here as they have no incidence on what follows).

The MRCA of the population at time $t$ is the same as the one of the population at time $t_0$ if and only if the ancestors $A_1$ and $A_2$ still have descendants in the population at time $t$. If $m$ is the number of ancestors at time $t_0$ of the population at time $t$, this means that some of these $m$ ancestors should be present in both groups of size $N_1$ and $N_2$ coming from $A_1$ and $A_2$ (see figure \ref{fig:arbrett0}). As the probabilities for each of the $m$'s to belong to the first or the second group are $x$ and $1-x$, the probability that both groups contains at least one of these $m$ ancestors is $1-(1-x)^m-x^m$.

If one introduces the probability $z_m(t-t_0)$ that the population at time $t$ has $m$ ancestors in the population at time $t_0<t$, the probability $P_\text{same}(t_0,t)$ is given by~:
\begin{equation}\label{def:Psame}
P_\text{same}(t_0,t) = \int_0^1 dx \sum_{m=2}^\infty z_m(t-t_0) \Big( 1-(1-x)^m - x^m \Big)
\end{equation}

The functions $z_m(t)$ are known \cite{derrida88}. They satisfy recursive equations:
the probability that the number of ancestors at $t_0<t$ of a population at $t$ is $m$
is the sum of the probability that this number is $m$ at
time $t_0+dt$ with no coalescence among these $m$ during $dt$
and of the probability that there are $m+1$ ancestors at $t_0+dt$ with a coalescence between $t_0$ and $t_0+dt$. Therefore the functions $z_m$ satisfy~:
\begin{equation}\label{zm:rec}
\frac{d}{d\tau} z_m(\tau) = c_{m+1} z_{m+1}(\tau) - c_m z_m(\tau)
\end{equation}
The function $z_1(\tau)$ is known as it is related to the distribution  (\ref{distristatT}) of the age $T$ of the MRCA~:
$$ \frac{d}{d\tau}z_1(\tau) = \rho_\text{st}(\tau) \quad\text{and}\quad z_1(0)=0$$
The solution of (\ref{zm:rec}) is \cite{derrida88}~:
\begin{equation}\label{def:zm}
z_m(\tau) = \sum_{p=m}^\infty
(-1)^{p+m}\frac{(2p-1)(m+p-2)!}{m!(m-1)!(p-m)!}e^{-c_p
\tau}\end{equation}

Using the normalization $\sum_{m=1}^\infty z_m(\tau)=1$ and the fact that $x$ is uniformly distributed between 0 and 1, one gets~:
\begin{eqnarray}
P_\text{same} (t_0,t) &=& 1-z_1(t-t_0) -\sum_{m=2}^\infty \frac{2}{m+1} z_m(t-t_0) \nonumber \\
&=& \sum_{p=2}^\infty (-1)^p (2p-1) e^{-c_p (t-t_0)} \Big[ 1 -2 \sum_{m=2}^p \frac{(-1)^m (m+p-2)!}{(m+1)! (m-1)! (p-m)!} \Big]  \label{Psame:decomp}
\end{eqnarray}
Using the identity~:
$$  \sum_{m=2}^p \frac{(-1)^m (m+p-2)!}{(m+1)! (m-1)! (p-m)!}  = \begin{cases} 1/3 & \text{if $p=2$} \\
1/2 & \text{if $p\geq 3$}
   \end{cases}$$
one can see that all the exponentials in (\ref{Psame:decomp}) vanish except the one for $p=2$ and one obtains~: 
\begin{equation}
P_\text{same}(t_0,t) = e^{-(t-t_0)}
\end{equation}
This shows that the delays $D_k$ between two successive jumps are distributed according to (\ref{D:exp})~:
$$ p_\text{delay}(D)= \frac{dP_\text{same}}{dt}(0,t=D)= e^{-D}$$

\subsection{The coalescence times $\tau_i$ as a Markov process}
\label{subsec:stoch}
%------------------------------------------------------------------------

\begin{figure*}
\begin{center}
\input{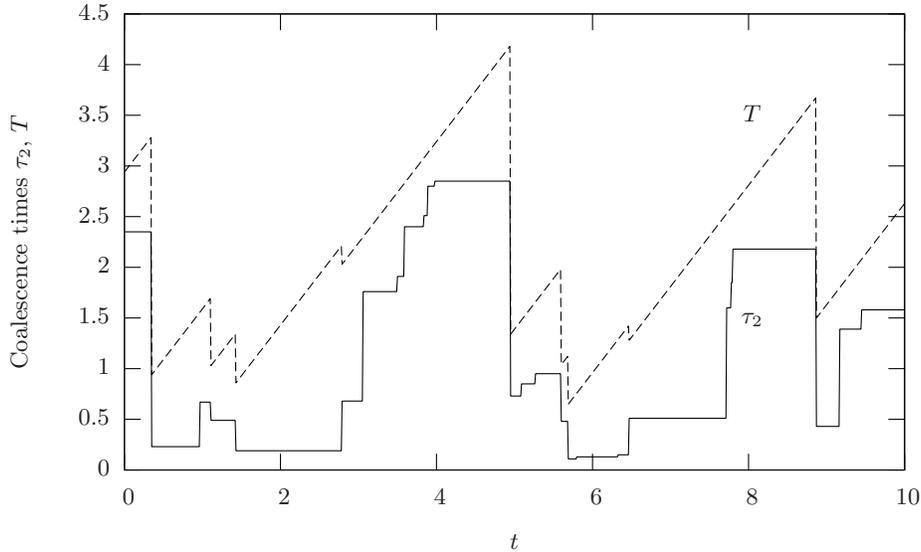}
\end{center}
\caption{Evolution of the delay $\tau_2$ between the two oldest coalescences at the top of the genealogical tree of a population of 100 individuals at time $t$. The dashed line corresponds to the age $T$ of the MRCA~: its shape is similar to figure \ref{fig:jumpfig}. The study of the dynamics of $\tau_2$ shows that, at random times depending on extinctions, the time $\tau_2$ either increases by $\tau_3$ or is reset to $\tau_3$, so that the new $\tau'_2$ is given either by $\tau_2+\tau_3$ or by $\tau_3$.}\label{fig:evoltau2}
\end{figure*}

%------------------------------------------------------------------

Figure \ref{fig:evoltau2} shows the stochastic dynamics of the coalescence time $\tau_2$. Actually, all the elementary times $\tau_i$ of figure \ref{fig:arbre} have similar dynamics. The coalescence times $\tau_i$ are the waiting times between two successive coalescences in a genealogy (see figure \ref{fig:arbre}) and evolve when extinctions of lineages occur. For example, if the lineage of $A_2$ in figure \ref{fig:arbre} gets extinct, then the new MRCA is $A_1$ and the new time $\tau'_2$ is the former $\tau_3$. This change implies a global shift $\tau'_i= \tau_{i+1}$ for $i\geq 2$. On the other hand, if the lineage of $A_1$ on the left gets extinct, the MRCA does not change but the $\tau_i$ become $\tau'_2=\tau_2+\tau_3$ and $\tau'_i= \tau_{i+1}$ for $i\geq 3$. 

More generally, one can consider the top of the genealogical tree of a population between the dates when the number of ancestors is 1 and $n$. In this part of the tree, there are $n-1$ coalescence times $\tau_2$, \ldots, $\tau_n$. The $n$ \emph{leaves} of the tree generate all the population in the present generation. The dynamics of the $\tau_i$ is controlled by the extinctions of the $n$ lineages coming from these $n$ ancestors~: whenever one of them gets extinct, some of the times $\tau_i$ topple. 

Actually, the observed dynamics of the $\tau_i$ can be described by the large $n$ limit of the following stochastic process~: either no extinction occurs and the times $\tau_i$ remain unchanged~:
\begin{equation} \tau_j(t+dt) = \tau_j(t) \quad \text{with probability $1-\frac{n(n-1)}{2}dt$} \end{equation}
or an extinction occurs and, with probability $p_idt$ for $2\leq i\leq n-1$, the times topple at rank $i$~:
\begin{equation}\label{def:cascade}\begin{cases}
\tau_j(t+dt) =  \tau_j(t) \quad \text{for $j<i$} \\
\tau_{i}(t+dt) = \tau_{i}(t)+\tau_{i+1}(t)  \\
\tau_{j}(t+dt) = \tau_{j+1}(t) \quad \text{for $i+1 \leq j \leq n-1$} \\
\tau_n(t+dt) = \epsilon_n(t)
\end{cases} \end{equation}

Moreover, with probability $p_1dt$, for $i=1$, all the times $\tau_j$ are shifted~:
\begin{equation}\label{def:cascadebis}\begin{cases}
\tau_j(t+dt) =  \tau_{j+1}(t) \quad \text{for $2\leq j\leq n-1$} \\
\tau_n(t+dt) = \epsilon_n(t)
\end{cases} \end{equation}

In appendix \ref{app:cascade}, we show that the toppling rates $p_i$ are given by~:
\begin{equation}
\label{prob:cascade}
p_i = i 
\end{equation} 

To determine the dynamics of $\tau_n$, we need to specify $\epsilon_n(t)$~: the $\epsilon_n(t)$ are random numbers uncorrelated in time which must have the same average as $\tau_n$~: $\langle \epsilon_n \rangle = \langle\tau_{n}\rangle=2/(n(n-1)) $. We will see however that, when $n$ is large, the precise form of the distribution of the $\epsilon_n$ plays no role as long as $\langle \epsilon_n\rangle=\langle \tau_{n}\rangle$. This feature can be understood because $c_n =n(n-1)/2$ goes to infinity when $n$ becomes large~; therefore the larger $n$ is, the more often the time $\tau_n$ is reset and a new $\epsilon_n$ enters the system~; in any time interval, the time $\tau_n$ is reset so many times with many independent $\epsilon_n$ entering the system that only $\langle\epsilon_n\rangle$ matters because of the law of large numbers.

The value of $\langle\epsilon_n\rangle$ can also be understood through the stationary conditions~: $\epsilon_n$ is added to the system with a rate $c_n$ whereas $\tau_2$ is removed with a rate $1$. The system can reach a stationary state only if $\langle\epsilon_n\rangle c_n = \langle \tau_2\rangle = 1$. Another consequence is that the total coalescence time $T(t)=\sum_{i=2}^n \tau_i$ increases on average by $c_n\langle \epsilon_n \rangle \Delta t=\Delta t$ during $\Delta t$ when no discontinuity occurs, in agreement with the slope $1$ observed in figures \ref{fig:jumpfig} and \ref{fig:evoltau2}.

These simple dynamics of the times $\tau_i$ allow one to determine all the statistical properties of $T(t)$~:  its correlations at different times, the distribution of its discontinuities $H_k$ and the distribution of the coalescence times $T$ right before a discontinuity.

First it is obvious that the distribution of delays between successive discontinuities of $T$ is exponential. The toppling dynamics (\ref{def:cascade}) imply also that, at a given time $t$, all the $\tau_i(t)$ are sums of times $\tau_j(0)$ with $j\leq i$ and of $\epsilon_n$'s. These sums do not overlap and thus, if the initial times $\tau_i(t)$ are not correlated at $t=0$, they remain uncorrelated at any later times. However, any $\tau_i$ depends only on previous $\tau_j$ with $i\leq j$ such that the only non zero correlations in this system are the $G_{i,j}(t)$ with $i\leq j$ defined as~:
\begin{equation}
\label{def:Gij}
G_{i,j}(t) = \langle \tau_i(t) \tau_j(0) \rangle -\langle \tau_i(t)\rangle \langle \tau_j(0)\rangle
\end{equation}

A consequence of (\ref{def:cascade}) and (\ref{def:cascadebis}) is that~:
\begin{equation}
\tau_i(t+dt) =\begin{cases}
 \tau_i(t) & \text{with probability $1-c_{i+1}dt$}\\
 \tau_i(t)+\tau_{i+1}(t) & \text{with probability $idt$}\\
 \tau_{i+1}(t) & \text{with probability $c_i dt$}
\end{cases} \label{tauicasc}
\end{equation}
Therefore, $G_{i,j}$ satisfies the following differential equation~:
\begin{equation}\label{rec:Gij}
\partial_t G_{i,j}(t) = -c_i G_{i,j}(t) + c_{i+1} G_{i+1,j}(t)
\end{equation}
The initial conditions correspond to the times $\tau_i$ generated according to the the stationary distribution (\ref{def:rhoi}) and thus one has $G_{i,j}(0) = \delta_{ij}/c_i^2$. We have seen that $G_{i,j}(t) = 0$ if $i>j+1$ and this gives immediately the solution of (\ref{rec:Gij}) for $i=j$~:
\begin{equation}
G_{j,j}(t) = \frac{1}{c_j^2} e^{-c_j t}
\end{equation}
More generally, the Laplace transform $\hat{G}_{i,j}(\lambda)=\int_0^\infty e^{-\lambda t} G_{i,j}(t)dt$ is given by the following product~:
\begin{equation}\label{res:Gij}
\hat{G}_{i,j}(\lambda) = \frac{1}{c_i c_j^2}\prod_{l=i}^j \frac{c_l}{\lambda+c_l}
\end{equation}

In particular, the correlation function of the total coalescence time $T(t)=\tau_2(t)+\tau_3(t)+\ldots$ can be written as~:
\begin{eqnarray}
\langle T(t) T(0) \rangle -\langle T(t)\rangle \langle T(0)\rangle &=& \sum_{i,j} \langle \tau_i(t) \tau_j(0)\rangle -\langle \tau_i(t)\rangle \langle \tau_j(0)\rangle \nonumber \\
&=&  \sum_{i=2}^\infty \sum_{j=2}^\infty G_{i,j}(t) \label{T0autocorr:casc}
\end{eqnarray}

In principle (\ref{res:Gij}, \ref{T0autocorr:casc}) allow one to extract the explicit expression of the autocorrelation function of $T$. We will describe later an alternative method to determine this explicit expression.

The dynamics (\ref{def:cascade}) gives also the statistical properties of the $\tau_i$ at the time of discontinuity. In particular, we are going to show that the distribution of the total coalescence time $T(t)$ right before a discontinuity is equal to the stationary distribution (\ref{distristatT}).

First, we remark from (\ref{def:cascade}) that a discontinuity occurs when $\tau_2$ is thrown out the system. More precisely the height $H_k$ is equal to this $\tau_2$ just before the discontinuity and the distribution of each $\tau_i$ just after the jump is the distribution of $\tau_{i+1}$ before the jump. Moreover, if the process is started at time 0 just after a discontinuity (i.e. we choose a discontinuity of $T$ as origin of time), one can introduce a variable $\eta(t)$ defined as~:
\begin{equation}
\label{def:eta}
\eta(t)  = \begin{cases}
1 & \text{before the next discontinuity of $T$}\\
0 & \text{after the next discontinuity of $T$}
           \end{cases}
\end{equation}
The dynamics (\ref{def:cascade}) implies that the average $\langle\eta\rangle$ decays exponentially as $\langle \eta(t) \rangle = e^{-t}$. 
The introduction of $\eta$ allows us to study what happens between discontinuities. In particular, the generating function $G^{(-)}(\lambda) =\langle e^{-\lambda T}\rangle_\text{before}$ of the coalescence time $T(t)$ right \emph{before} a discontinuity takes the form~:
\begin{equation}\label{Tdiscont:def}
G^{(-)}(\lambda)=\langle e^{-\lambda T}\rangle_\text{before} = \int_0^\infty \langle \eta(t) e^{-\lambda T(t)}\rangle dt
\end{equation}
From (\ref{def:cascade}),the correlation function $\langle \eta(t) e^{-\lambda T(t)}\rangle$ satisfies~:
$$ \partial_t\langle \eta(t) e^{-\lambda T(t)}\rangle = - c_n \langle \eta(t) e^{-\lambda T(t)}\rangle + (c_n-1) \langle \eta(t) e^{-\lambda T(t)}\rangle \langle e^{-\lambda\epsilon_n} \rangle$$
Integrating over $t$, one gets for $G^{(-)}(\lambda)$ from (\ref{Tdiscont:def})~:
\begin{equation}
\label{res:G-}
G^{(-)}(\lambda) = \frac{ 1}{ c_n - (c_n-1)\langle e^{-\lambda \epsilon_n}\rangle}G^{(+)}(\lambda)
\end{equation}
where $G^{(+)}(\lambda) =\langle e^{-\lambda T}\rangle_\text{after}$ is the generating function of the total coalescence time right \emph{after} the discontinuity.

On the other hand, the stationary distribution can also be written in terms of $G^{(+)}$. The generating function $ \langle e^{-\lambda T(t)}\rangle$ of $T(t)$ satisfies~:
\begin{equation}
\langle e^{-\lambda (T(t)+dt)} \rangle = (1-c_n dt) \langle e^{-\lambda T(t)} \rangle + dt \langle e^{-\lambda T(t)} \rangle_\text{after} + (c_n-1) \langle e^{-\lambda T(t)} \rangle \langle e^{-\lambda\epsilon_n}\rangle
\end{equation}
Thus, the stationary generating function is given by~:
\begin{equation}
\label{res:Gstat}
G_\text{st} = \langle e^{-\lambda T(t)} \rangle_\text{st} = \frac{1}{c_n-(c_n-1)\langle e^{-\lambda\epsilon_n}\rangle} G^{(+)}(\lambda)
\end{equation}

Comparing (\ref{res:G-}) and (\ref{res:Gstat}), we see that~:
\begin{equation}
\label{G-Gst}
G^{(-)}(\lambda)= \langle e^{-\lambda T}\rangle_\text{before}  = \langle e^{-\lambda T(t)} \rangle_\text{st} = G_\text{st}(\lambda)
\end{equation}

This result, which we checked in our simulations, looks paradoxical~: although $T(t)$ reaches a local maximum when the MRCA changes, the distribution of $T$ at these local maxima is the same as the distribution of $T(t)$ over the whole range of time. In fact, one can show by similar calculations that the same is true for all the $\tau_i$'s~: their distributions right before a discontinuity of $T$ are the same as the stationary ones. 
The case of $\tau_2$ explains the properties of the drops $H_k$ at the discontinuities of $T$, since the value of $H_k$ is the value of $\tau_2$ just before the discontinuity. Their distribution is exponential~:
\begin{equation}
\label{H:exp}
p_\text{height}(H) = e^{-H}
\end{equation}
which is in agreement with the data of figure \ref{fig:jumpstatfig}. Moreover, the $H_k$ are not correlated in agreement with (\ref{Hautocorr}), as if $H_k=\tau_2$, then $H_{k+1}$ is made of some $\tau_j$'s with $j\geq 3$ at the time of the previous discontinuity.

One also sees from (\ref{def:cascade}) that, just after the discontinuity, $\tau_i$ is replaced by $\tau_{i+1}$ just before the discontinuity, which was distributed according to the stationary distribution (\ref{def:rhoi}). Thus the distribution $G^{(+)}$ should be given by a formula similar to (\ref{qpopnoinfo}) starting only at $l=3$. The comparison with (\ref{res:Gstat}) implies that the factor $1/(c_n-(c_n-1)\langle e^{-\lambda\epsilon_n}\rangle)$ should become $1/(\lambda+1)$ for the large $n$ limit. This is easily checked as $\epsilon_n\sim1/n^2$ and for large $n$~:
$$ \langle e^{-\lambda\epsilon_n}\rangle = 1-\lambda\langle \epsilon_n\rangle+o(1/n^2)$$ 
This in particular shows that for large $n$, only the average of $\langle \epsilon_n \rangle$ matters.

The analytical value of (\ref{DHcorr}) can also be obtained using the toppling dynamics of the $\tau_i$. Using the variable $\eta(t)$ defined in (\ref{def:eta}), the delay $D_k$ is the time at which $\eta(t)$ goes to zero and the height $H_k$ is the time $\tau_2$ right before the drop. The correlation coefficient is given by~:
$$ \langle D_k H_k\rangle = \int_0^\infty t \langle \eta(t)\tau_2(t)\rangle dt $$
This suggests to consider the functions $\psi_i(\lambda) = \int_0^\infty e^{-\lambda t}\langle \eta(t) \tau_i(t) \rangle dt$, as the correlation coefficient is $\langle D H\rangle = -d\psi_2/d\lambda$ for $\lambda=0$. The coefficients $\langle \eta(t) \tau_i(t) \rangle$ satisfy the following differential equation derived from (\ref{def:cascade}) and (\ref{tauicasc})~:
\begin{equation}
\frac{d}{dt}\langle \eta(t) \tau_i(t) \rangle = -c_i \langle \eta(t) \tau_i(t) \rangle + (c_{i+1}-1) \langle \eta(t) \tau_{i+1}(t) \rangle
\end{equation}
At $t=0$, $\eta(t)$ is equal to $1$ and $\tau_i$ is distributed according to $\rho_{i+1}$ given by (\ref{def:rhoi}), since we saw in (\ref{def:cascadebis}) that there is a global shift of the $\tau_i$'s at each discontinuity of $T$. This implies that the functions $\psi_i$ satisfy the following recursion~:
\begin{equation}
\label{rec:psiDH}
\lambda \psi_i(\lambda) -\frac{1}{c_{i+1}} = - c_i \psi_i(\lambda) + (c_{i+1}-1) \psi_{i+1}(\lambda)
\end{equation}
As we need the first derivative of $\psi_2$ in zero, we can expand $\psi_i$ in powers of $\lambda$~:
\begin{equation}
\psi(\lambda) = \frac{2}{i(i+1)}\Big( u_i - \lambda v_i + O(\lambda^2)\Big)
\end{equation}
The coefficients $u_i$ and $v_i$ satisfy the following simple recursion derived from (\ref{rec:psiDH})~:
\begin{subequations}
\begin{eqnarray}
u_i &=& \frac{2}{i(i-1)} + u_{i+1} \label{rec:ui}\\
v_i &=& \frac{u_i}{i(i-1)} + v_{i+1} \label{rec:vi}
\end{eqnarray}
\end{subequations}
The term $\psi_{n+1}$ is linked to the boundary condition $\epsilon_n$ and one has $\psi_{n+1}(\lambda) = \langle \epsilon_n\rangle/(1+\lambda)$ so that $u_{n+1} = \langle \epsilon_n\rangle(n+1)(n+2)/2$ and $v_{n+1} = \langle \epsilon_n\rangle(n+1)(n+2)/2$ (which are not negligible when $n\rightarrow\infty$).
Equations (\ref{rec:ui}) and (\ref{rec:vi}) give simple summation formulas for $u_i$ and $v_i$~:
\begin{subequations}
\begin{eqnarray}
u_i &=& \sum_{j=i}^n \frac{2}{j(j-1)} + \langle \epsilon_n\rangle(n+1)(n+2)/2 \underset{n\rightarrow\infty}{\longrightarrow} \frac{2}{i-1}+1\\
v_i &=& \sum_{j=1}^n \frac{u_j}{j(j-1)} + \langle \epsilon_n\rangle(n+1)(n+2)/2 \underset{n\rightarrow\infty}{\longrightarrow} \frac{2\pi^2}{3}+1-\frac{2}{i-1}
\end{eqnarray}
\end{subequations}
Finally, the expansion of $\psi_2$ around $\lambda=0$ gives the following correlation coefficient in good agreement with the measured value (\ref{DHcorr})~:
\begin{equation}\label{DHcorr:theo}
\langle D_k H_k\rangle - \langle D_k\rangle \langle H_k\rangle = \frac{2\pi^2}{9}-\frac{4}{3} \simeq 0.8599\ldots
\end{equation}

\subsection{Correlation functions of the coalescence times between few individuals}

Consider a pair $(i,j)$ of individuals at generation $t$. One can define the time $T^{(i,j)}(t)$ to find their first common ancestor (i.e. $NT^{(i,j)}(t)$ is the number of generations to reach their first common ancestor). Similarly, one may consider three individuals $(i,j,k)$ at generation $t$ and define the time $T^{(i,j,k)}(t)$ to find their first common ancestor. One can average these times over the whole population~:
\begin{eqnarray}\label{def:Tiaver}
T_2(t) &=& \frac{1}{N^2} \sum_{i,j} T^{(i,j)}(t) \\
T_3(t) &=& \frac{1}{N^3} \sum_{i,j,k} T^{(i,j,k)}(t) \\
\end{eqnarray}

Figure \ref{fig:jumpfig} shows the stochastic evolution of
these averages $T_2(t)$ and $T_3(t)$. We are now going to determine the correlation functions of these times (in order to avoid confusion, we will use lower case letters $t$ for the usual time (oriented towards the future) and upper case letter $T$ for ages (i.e. oriented towards the past)).

To understand the correlations of $T_2(t)$ and $T_3(t)$, let us look at two individuals
$i$ and $j$ at generation $t$ and two individuals $k$ and $l$ at
generation $0$. Their coalescence times are defined as $T^{(i,j)}(t)$ and $T^{(k,l)}(0)$. 
There are two possibilities~: 
\begin{itemize}
\item either $T^{(i,j)}$ is smaller than $t$ and the coalescence times $T^{(i,j)}(t)$
and $T^{(k,l)}(0)$ are independent, 
\item $T^{(i,j)}$ is larger than $t$ and the entanglement between
lineages creates a correlation between $T^{(k,l)}(0)$ and $T^{(i,j)}-t$.
In the large population limit $N\rightarrow\infty$, the
probability that the ancestors of $i$ and $j$ are $k$ or $l$ goes
to 0 as $1/N$~; thus, in the second case, the quantity
$\langle(T^{(i,j)}(t)-t)T^{(k,l)}(0)\rangle$ is the average of the product
of the coalescence times of two distinct pairs of individuals at the generation 0. 
\end{itemize}

As a result, the average over the population $T_2(t)$ of
the coalescences times of two individuals $T_2^{(i,j)}$ satisfies~:
\begin{equation}
\langle T_2(t) T_2(0) \rangle \underset{N\rightarrow\infty}{=} 
\int_0^t\rho_2(\tau_2)\tau_2 d\tau_2 \times \langle T_2(0)\rangle
 +e^{-t} \Big( t\langle T_2(0)\rangle + \langle T^{(1,2)}(0)
T^{(3,4)}(0)\rangle\Big)
\end{equation}

The coefficient $\langle T^{(1,2)}(0) T^{(3,4)}(0)\rangle$ can be calculated by looking at the genealogy of only four individuals. Following appendix \ref{measgenstruct}, the coalescence times $T^{(1,2)}(0)$ and $T^{(3,4)}(0)$ are sums of the three elementary coalescences times $\tau_2$, $\tau_3$ and $\tau_4$. These decompositions are shown in figure \ref{decompTijTkl}.  Averaging over the tree structures and the times $\tau_i$ leads to~:
\begin{equation}\label{res:T2autocorr}
\langle T_2(t) T_2(0)\rangle  = 1+\frac{2}{9}e^{-t}\end{equation}
A similar calculation of the coalescence time of three individuals leads to~:
\begin{equation}\label{res:T3autocorr}
\langle T_3(t) T_3(0)\rangle  = \frac{16}{9} + \frac{29}{60}e^{-t}
-\frac{13}{900} e^{-3t}
\end{equation}

\begin{figure*}
\begin{center}
\begin{fmffile}{arbre4}
\begin{tabular}{|c|c|c|c|c|}
\hline Tree structure  &  $T^{(1,2)}$ &  $T^{(3,4)}$ & Symmetry factor \\
\hline
\begin{arbre}
\fmfstraight \fmftop{i1} \fmfstraight \fmfbottom{a1,a2,b1,b2}
\fmflabel{$1$}{a1} \fmflabel{$2$}{a2} \fmflabel{$3$}{b1} \fmflabel{$4$}{b2}
\fmf{plain}{i1,a}\fmf{plain}{i1,b} \fmf{plain}{a,a1}
\fmf{plain}{a,a2} \fmf{plain}{b,bb1} \fmf{plain}{b,bb2}
\fmf{plain}{bb1,bbb1} \fmf{plain}{bb2,bbb2} \fmf{plain}{bbb1,b1}
\fmf{plain}{bbb2,b2}
\end{arbre}
&  $\tau_4$ & $\tau_4+\tau_3$ & $\frac{2}{18}$ \\
\hline
\begin{arbre}
\fmfstraight \fmftop{i1} \fmfstraight \fmfbottom{a1,a2,b1,b2}
\fmflabel{$1$}{a1} \fmflabel{$3$}{a2} \fmflabel{$2$}{b1} \fmflabel{$4$}{b2}
\fmf{plain}{i1,a}\fmf{plain}{i1,b} \fmf{plain}{a,a1}
\fmf{plain}{a,a2} \fmf{plain}{b,bb1} \fmf{plain}{b,bb2}
\fmf{plain}{bb1,bbb1} \fmf{plain}{bb2,bbb2} \fmf{plain}{bbb1,b1}
\fmf{plain}{bbb2,b2}
\end{arbre}
& $\tau_4+\tau_3+\tau_2$ & $\tau_4+\tau_3+\tau_2$ & $\frac{4}{18}$ \\
\hline
\begin{arbre}
\fmfstraight \fmftop{i1} \fmfstraight \fmfbottom{a1,a2,b1,b2}
\fmflabel{$1$}{a1} \fmflabel{$2$}{a2} \fmflabel{$3$}{b1} \fmflabel{$4$}{b2}
\fmf{plain}{i1,b2}\fmf{plain}{i1,a} \fmf{plain}{a,b1}
\fmf{plain}{a,b} \fmf{plain}{b,a1} \fmf{plain}{b,a2}
\end{arbre}
& $\tau_4$ & $\tau_4+\tau_3+\tau_2$ & $\frac{4}{18}$ \\
\hline
\begin{arbre}
\fmfstraight \fmftop{i1} \fmfstraight \fmfbottom{a1,a2,b1,b2}
\fmflabel{$1$}{a1} \fmflabel{$3$}{a2} \fmflabel{$2$}{b1} \fmflabel{$4$}{b2}
\fmf{plain}{i1,b2}\fmf{plain}{i1,a} \fmf{plain}{a,b1}
\fmf{plain}{a,b} \fmf{plain}{b,a1} \fmf{plain}{b,a2}
\end{arbre}
& $\tau_4+\tau_3$ & $\tau_4+\tau_3+\tau_2$ & $\frac{8}{18}$ \\
\hline
\end{tabular}
\end{fmffile}
\end{center}
\caption{\label{decompTijTkl} Genealogical trees of four individuals $1$, $2$, $3$ and $4$  and the corresponding decomposition of the coalescence times of individuals $1$ and $2$ on one hand, and $3$ and $4$ on the other hand. Up to symmetries, there are only these five types of decomposition~: any other tree leads to the same type of decomposition (up to permutations of the labels or of the roles of $(1,2)$ and $(3,4)$). The symmetry factors count these relabellings.}
\end{figure*}

More generally the correlation functions of coalescence times
$T_m$ would be a linear combination of $e^{-c_p t}$ weighted by
coefficients. The calculation of the correlation function of the $T_m$ becomes however more and more complicated with increasing $m$. We have only been able to determine the correlation
function $\langle T(t)T(0)\rangle - \langle T(t)\rangle\langle
T(0)\rangle$ of the coalescence time of the whole population
represented in figure \ref{fig:jumpfig}. As for $T_2$, one has to consider two cases~: either the MRCA of the population at $t$ is reached between $0$ and $t$ so that $T(t)<t$, or the number of ancestors at $0$
is $m\geq 2$ so that $T(t)= t+T_m(0)> t$. If $z_m(\tau)$ is the
probability (\ref{def:zm}) that the number of ancestors of the population after a
duration $\tau$ in the past is $m$ , we have the following
decomposition~:
\begin{equation}\label{T:decomp}
\langle T(t)T(0)\rangle = \int_0^t \tau z'_1(\tau)d\tau \times
\langle T(0)\rangle + \sum_{m\geq 2} z_m(t) \big\langle (t+T_m(0)) T(0)\big\rangle
\end{equation}
where $z'_1(\tau) =\rho_\text{st}(\tau) =
\text{Prob}(T=\tau)$ is the probability that the MRCA is reached
at $\tau$.

The coefficients $\langle T_m(0) T(0)\rangle$ can be decomposed in
a tree-depending combination of the elementary times $\tau_i$ (see
section \ref{subsec:intro} and appendix \ref{measgenstruct})
with~:
\begin{equation} T_m(0) = \sum_{i=q+1}^\infty \tau_i
\quad\text{and}\quad T(0) = \sum_{i=2}^\infty \tau_i
\end{equation}
where $q$ is the number of ancestors left from the whole population
when the subgroup of size $m$ has just coalesced. If
$a_{m,\infty}(q)$ is the probability distribution of $q$, then one
has~:
\begin{equation}\label{TmT0:decomp}
\langle T_m(0) T(0)\rangle = \sum_{q=1}^\infty a_{m,\infty}(q)
\sum_{i=q+1}^\infty \sum_{j=2}^\infty \langle \tau_i\tau_j\rangle = \sum_{i=2}^\infty \Big[ \sum_{j=2}^\infty \langle \tau_i\tau_j\rangle \Big] \Big[ \sum_{q=1}^{i-1} a_{m,\infty}(q) \Big]
\end{equation} where the $\tau_i$ are independent random variables
with exponential distribution (\ref{def:rhoi}).

Let us define $a_{m,n}(q)$ as the probability that
the number of ancestors of a group of size $m+n$ is $q$ at the time when the first
subgroup of $m$ individuals has just coalesced into a single ancestor. Writing all the
possibilities for the first coalescence of the group of size $m+n$
leads to the following recursive equation~:
\begin{equation}\label{amnq:rec}
a_{m,n}(q) = \frac{ c_m a_{m-1,n}(q) +(c_n+nm)
a_{m,n-1}(q)}{c_{n+m}}
\end{equation}
The boundary conditions are the probability that coalescences
occur only among the first $m$ if $q=n+1$~:
$$a_{m,n}(n+1)=\prod_{i=2}^m \frac{c_i}{c_{i+n}}= \frac{m!(m-1)! n! (n+1)!}{(m+n)!(m+n-1)!}$$
and the probability for $m=2$ that two individuals coalesce once the $n$ others are reduced to $q$~: $$a_{2,n}(q)=
\frac{c_2}{c_{q+1}}\prod_{j=q}^n\frac{c_j+2j}{c_{j+2}} =
\frac{2}{(q+2)(q+1)}\frac{n+3}{n+1}$$
With these boundary conditions, one gets for the solution of (\ref{amnq:rec})~:
\begin{equation}\label{amnq:general}
a_{m,n}(q)=\frac{ m!n!q!(m+n-q-1)!
(m-1)(m+n+1)}{(q+m)!(m+n-1)!(n-q+1)!}
\end{equation}
and in the limit $n\rightarrow\infty$~:
\begin{equation}\label{aminftyq} a_{m,\infty}(q) = \frac{q!m!(m-1)}{(m+q)!}
\end{equation}
One can ckeck easily that~:
\begin{equation}\label{sum:amq}
\sum_{q=1}^{i-1} a_{m,\infty}(q) = 1 -\frac{i! m!}{(m+i-1)!}
\end{equation}
Moreover, using (\ref{def:rhoi}), the correlation between $\tau_i$ and $\tau_j$ is~:
\begin{equation}\label{correltaui_tauj}
\langle \tau_i\tau_j \rangle = \frac{1}{c_i c_j} (1+\delta_{ij})
\end{equation}
Using (\ref{correltaui_tauj}) and (\ref{sum:amq}), the permutation of the sums in (\ref{TmT0:decomp}) gives the correlation coefficients $\langle T_m(0) T(0)\rangle$~:
\begin{eqnarray*}
\langle T_m(0) T(0) \rangle
&=& \sum_{i=2}^\infty \Big[ \frac{1}{c_i^2}+\frac{2}{c_i} \Big] \Big[ 1 -\frac{i!m!}{(m+1-i)!} \Big] \\
&=& \sum_{i=2}^\infty \Big( \frac{1}{c_i^2}+\frac{2}{c_i} \Big) - 4\sum_{i=2}^\infty \frac{  m! (i-2)!}{(m+i-1)!} -\sum_{i=2}^\infty \frac{ 4 m! i! }{i^2(i-1)^2 (m+i-1)!}
\end{eqnarray*}

\begin{figure}[!ht]
  \begin{center}
  \input{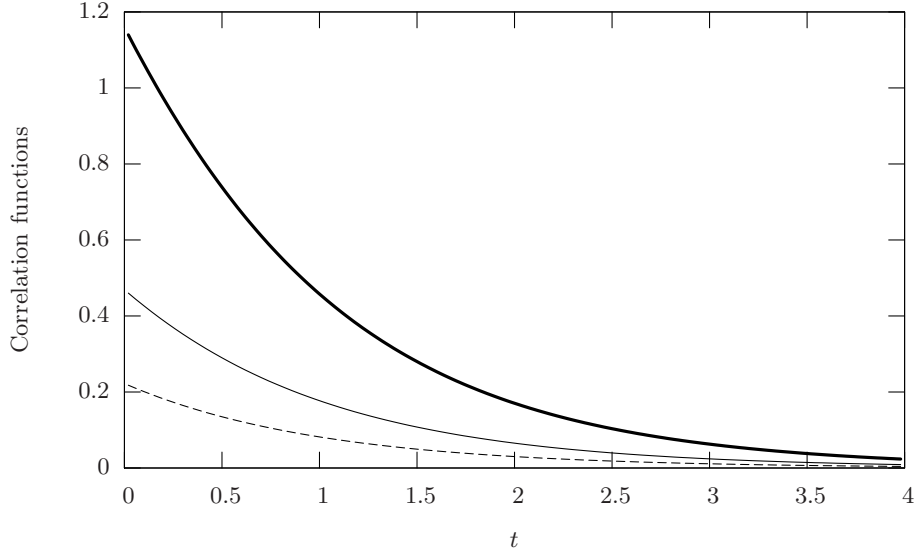}
  \end{center}
\caption{Correlation functions of the average over the population $T_2=1/(N^2)\sum_{i, j} T^{(i,j)}$ of the age of the MRCA for $n=2$ individual (dashed), of the average $T_3=1/(N^3)\sum_{i,j,k} T^{(i,j,k)}$ of the age of the MRCA of $n=3$ individuals (thin) and of the age $T(t)$ of the MRCA of the whole population (thick).}
\label{fig:Tautocorr}
\end{figure}

The calculations of the first two sums give~:
\begin{eqnarray*}
\sum_{i=2}^\infty \Big( \frac{1}{c_i^2}+\frac{2}{c_i} \Big) &=& \frac{4\pi^2}{3} - 8 \\
\sum_{i=2}^\infty  \frac{  m! (i-2)!}{(m+i-1)!} &=& \frac{1}{m}
\end{eqnarray*}
and $\langle T_m(0) T(0)\rangle$ becomes~:
\begin{equation}\label{TmT0:res}
\langle T_m(0) T(0) \rangle = \frac{4\pi^2}{3}-8 - \frac{4}{m} -  \sum_{i=2}^\infty \frac{4 m! i!}{i^2(i-1)^2 (m+i-1)!}
\end{equation}
Finally, using the normalisation $\sum_{m=1}^\infty z_m(t)=1$ and the fact that $\langle T(0)\rangle =2$, the integration of the first term of (\ref{T:decomp}) leads to~: 
$$ \langle T(t)T(0) \rangle = 2\int_0^t(1-z_1(t))dt + \sum_{m=2}^\infty \langle T_m(0)T(0) \rangle z_m(t) $$
By multiplying (\ref{zm:rec}) by $1/m$ and summing over $m$, one gets~:
\begin{equation}\label{propzm:2}
\frac{d}{d\tau}\Big( \sum_{m=2}^\infty \frac{1}{m} z_m(\tau) \Big) = -\frac{d}{d\tau}z_1(\tau) +\frac{1}{2} (1-z_1(\tau))
\end{equation}
Since the sum $\sum_{m=2}^\infty z_m(\tau)/m$ must vanish for large $\tau$, the solution of (\ref{propzm:2}) is~:
\begin{equation}\label{zm:formula1}\sum_{m=2}^\infty \frac{1}{m} z_m(t) = (1-z_1(t)) -\frac{1}{2}\int_t^\infty(1-z_1(t)) dt \end{equation}
Using (\ref{TmT0:res}) and (\ref{zm:formula1}) one gets~:
$$ \langle T(t)T(0) \rangle = 4+ \big(\frac{4\pi^2}{3}-12\Big)(1-z_1(t)) - \sum_{m=2}^\infty \sum_{i=2}^\infty \frac{4m!i!}{i^2(i-1)^2(m+i-1)!} z_m(t) $$
If one collects the exponential terms $e^{-c_p t}$ using (\ref{def:zm}), the correlation function takes the following form~:
\begin{equation}\label{T:autocorr}
\langle T(t)T(0)\rangle - \langle T(t)\rangle\langle T(0)\rangle =
\sum_{p=2}^\infty (-1)^p (2p-1) A_p e^{-c_p t}
\end{equation}
with coefficients $A_p$ given by~:
\begin{equation}
A_p = \frac{4\pi^2}{3}-12 - \sum_{i=2}^\infty \sum_{m=2}^p \frac{4 i! (-1)^m (m+p-2)!}{i^2 (i-1)^2(m+i-1)!(m-1)!(p-m)!}
\end{equation}
One can show that the sum over $m$ is given by~:
$$ \sum_{m=2}^p \frac{(m+p-2)!(-1)^m}{(m+i-1)!(m-1)!(p-m)!} = \frac{1}{i!}- 
\begin{cases}
0 & \text{for $p>i$} \\
\frac{(i-1)!}{(p+i-1)!(i-p)!} & \text{for $i\geq p$} 
\end{cases}$$
This identity gives finally the coefficient $A_p$~:
\begin{equation}
\label{T:coeffAp}
A_p = 4 \sum_{i=p}^\infty \frac{(i-2)!^2}{i (i+p-1)!(i-p)!}
\end{equation}
One can notice that $A_p{\longrightarrow}0$ when $p\rightarrow\infty$. The correlation functions $\langle T_2(t) T_2(0)\rangle- \langle T_2(t)\rangle\langle T_2(0)\rangle$, $\langle T_3(t) T_3(0)\rangle-\langle T_3(t) \rangle \langle T_3(0)\rangle$ and $\langle T(t) T(0)\rangle -\langle T(t) \rangle \langle T(0)\rangle$ are shown in figure \ref{fig:Tautocorr}.

By a calculation not shown here, one can check that expressions (\ref{T:autocorr},\ref{T:coeffAp}) coincide with the one obtained from (\ref{T0autocorr:casc}) and this confirms the validity of the Markov process defined in (\ref{def:cascade}) and (\ref{def:cascadebis}).

\section{Correlation between the coalescence time and the genomic diversity}\label{corrdiv}

So far we only considered the statistical properties of the coalescence times along the tree. We are going to study now how these times are correlated to the genetic diversity.

%---------------------------------------------------------------

The genetic diversity can be measured by different quantities according to the model one considers (see for example Tajima's estimator for the infinite site model \cite{tajima}). We consider here the case of an infinite number of alleles~: any mutation creates a new allele which has never occured before. Thus, for two individuals chosen at random in the population, there are only two possibilities~: either they have the same allele or they have different ones. Now we want to calculate the average age of the MRCA, conditioned on the fact that the two individuals chosen at random have (or not) the same genome.

More generally, the population is divided into groups of individuals sharing the same genome, whose sizes characterize the genetic diversity of the population. The determination of the distribution of the age of the MRCA, given
the size of these groups, is a difficult problem that we could not solve. Here we address a simpler version of this problem~: suppose we have some information about the genes of a few individuals chosen at random in the population; what can be said about the age of the MRCA~?

In the present case, we consider a group of $n\ll N$ individuals
and we suppose that the first $m$ of them have identical genomes. Of
course, the $n-m$ others may have the same genome or different
ones~: we suppose that we have no information about them. Knowing this partial
information about the present generation, we look at the
coalescence time of the whole group of $n$ individuals. 

We first look at the probability distribution $p_{m,n}(T_n)$
of observing a group of size $n$ whose coalescence time is equal
to $T_n$ and in which the first $m$ individuals have the same
genome. The coalescence time $T_n$ of such a group of size
$n$ is the coalescence time of their parents at the previous
generation plus one generation . The group of the parents is a group of size
$n'\leq n$. At first order in $1/N$, the only possible events which may occur are
a coalescence ($n'=n-1$) or a mutation ($n'=n$). The probability
of a coalescence among the first $m$ individuals is
$c_m/N=m(m-1)/2N$~; in this case, the probability distribution of
the coalescence time of the parents is $p_{m-1,n}$. For other
coalescences (probability $(c_n-c_m)/N$), it is $p_{m,n-1}$.
Moreover, no mutation must affect the first $m$ individuals.
Consequently, the probability distribution $p_{m,n}(T)$ satisfies
the following recursive equation~:
\begin{equation}\label{1group:p_rec}
\frac{d}{dT} p_{m,n}(T) =  c_m p_{m-1,n-1}(T)+\Big(c_n-c_m\Big) p_{m,n-1}(T) 
-\Big(c_n+m\theta\Big) p_{m,n}(T) 
\end{equation}
where the $c_n$ are the binomial coefficient (\ref{def:ci}).

For $m=1$, the distribution $p_{1,n}$ is just the stationary distribution of $T_n$ related to (\ref{qknoinfo}). For $n=m$, $p_{m,m}$ is the distribution of the coalescence time of a group of $m$ individuals with the same genome. Its Laplace transform is \cite{hein}~:
\begin{equation}
\hat{p}_{m,m}(s) = \int_0^\infty p_{m,m}(t)e^{-st} dt = \prod_{i=2}^m \frac{c_i}{s+c_i+i\theta}
\end{equation}

%------------------------------------------------------------------

The general solution of (\ref{1group:p_rec}), which we will give below in (\ref{psi_nm}), is difficult to handle in general. Let us consider first the simple case $m=2$ and define the parameter $Y$ related to the genomic
diversity as~:
\begin{equation}\label{def:Y}
 Y = \frac{1}{N(N-1)}\sum_{i\neq j} \delta_{g(i),g(j)}
\end{equation}
where $g(i)$ is the genome of the individual $i$. 
$Y$ doesn't count the number of differences between two sequences (as does Tajima's estimator \cite{tajima}) since we do not suppose any information about the structure of the genome but just detects whether at least one mutation has occurred or not and can be interpreted as the fraction of pairs of individuals having the same genome. When $Y$ is close to $1$, the population is very homogeneous and all the individuals have very similar
genomes whereas $Y$ close to $0$ corresponds to a population where the genetic diversity is very large. From the definition of $p_{m,n}$, one gets~:
\begin{equation}\label{link_q2_Y}
\hat{p}_{2,\infty}(s) = \langle Ye^{-sT}\rangle
\end{equation}
where $\hat{p}_{2,\infty}(s)$ is the limit for large $n$ of the generating functions $\hat{p}_{2,n}(s)$ that satisfy a recursion directly deduced from (\ref{1group:p_rec})~:
\begin{equation}\label{rec:p2ns}
\hat{p}_{2,n}(s) = \frac{1}{s+c_n+2\theta}	\Big( \hat{p}_{n-1}(s) + (c_n-1) \hat{p}_{2,n-1}(s) \Big)
\end{equation}
where $\hat{p}_{n}(s)=\langle e^{-sT_n} \rangle$ is the generating function with no information (\ref{qknoinfo}). The solution of (\ref{rec:p2ns})  (which is a particular case of the general solution (\ref{psi_nm}) given below) is~:
\begin{equation} \label{hatp2infty}
\hat{p}_{2,\infty}(s) = \langle Y e^{-sT}\rangle=\sum_{q=1}^\infty\frac{2}{(q+2)(q+1)}
\prod_{i=2}^{q}
\frac{c_i}{s+c_i}\prod_{j=q+1}^\infty\frac{c_j}{s+c_j+2\theta}\end{equation}

It allows one to determine the distribution of the coalescence time of the whole population, conditioned on the fact that two individuals chosen at random have the same genome. 
Moreover, successive derivations of (\ref{hatp2infty}) in $s=0$ give all the correlation coefficients $\langle Y T^k\rangle$. These coefficients measure how $Y$ is an estimator well adapted to the determination of the age of the MRCA $T$. The following computation focuses on the properties of the average coalescence time $\langle T | \text{2 id.}\rangle$ knowing that two individuals chosen at random have the same genome.

The average coalescence time $T_n$ of $n$ individuals conditioned on the fact that two individuals chosen at random among these $n$ have the same genome can also be obtained from (\ref{1group:p_rec}). The Laplace transform $\hat{p}_{m,n}(s)=\int_0^\infty
e^{-sT}p_{m,n}(T)dT$ for $s=0$ gives the probability that the
first $m$ individuals of the group of size $n$ have the same
genome (see (\ref{Pgroups})). Thus the normalized quantity $\hat{p}_{m,n}(s)/\hat{p}_{m,n}(0)$ is the generating functions of the coalescence time of $n$ individuals conditioned on the fact that $m$ individuals chosen at random among them have the same genome. For $m=2$, one has $\hat{p}_{2,n}(0)=1/(1+2\theta)$. The average conditioned time is the derivative of $\hat{p}_{m,n}(s)/\hat{p}_{m,n}(0)$ for $s=0$~:
$$ u_n(\theta) =\langle T_n | \text{ 2 id.} \rangle = -(1+2\theta)\frac{d}{ds}\hat{p}_{2,n}(s)\Big|_{s=0}$$
By taking the derivative of (\ref{rec:p2ns}) one gets~:
\begin{equation}
\label{rec:un}
u_n(\theta) = \frac{1}{c_n+2\theta} \Big( 1+ 2(1+2\theta)\frac{n-2}{n-1} + (c_n-1)u_{n-1}(\theta)\Big) 
\end{equation}
The initial condition is given by the coalescence time of 2 individuals with the same genome~:
$$ u_2(\theta)= \frac{1}{1+2\theta}$$
The general solution of (\ref{rec:un}) is given by~:
\begin{equation}\label{res:un}
u_n(\theta) =  \frac{2(n-2)}{n-1} + \frac{1}{1+2\theta} + \sum_{p=3}^n \frac{(-1)^p}{c_p+2\theta} \frac{ (n+1)!(n-2)!}{(n+p-1)!(n-p)!} \frac{(2p-1)(p+1)(p-2)}{2}
\end{equation}

If $\theta=0$, all the individuals have the same genome and the value of $u_n(\theta)$ for $\theta=0$ is just $2(n-1)/n$ as given by (\ref{tmoy_unknown}). The large $n$ limit of (\ref{res:un}) (performed by considering $u_n(\theta)-u_n(0)$ to regularize the series) leads for the average coalescence time $\langle T | \text{2 id.} \rangle$ of a whole population conditioned on the fact that two individuals chosen at random have identical genomes to~:
\begin{equation}
\label{res:Tconditionned2}
\langle T | \text{2 id.} \rangle =  1+\frac{1}{1+2\theta} - 2\theta \sum_{p=3}^\infty \frac{(2p-1)(p+1)(p-2)}{2} \frac{(-1)^p}{c_p(c_p+2\theta)}
\end{equation}

\begin{figure*}[t]
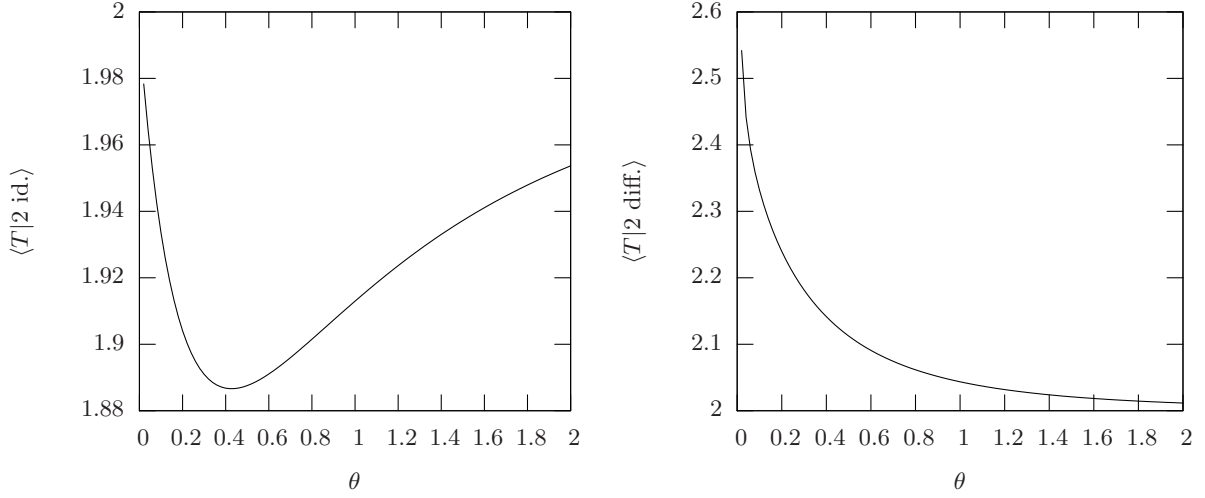

\begin{center}
\begin{tabular}{p{8cm}p{8cm}}
\input{t2id.tex} & \input{t2diff.tex} \\
\end{tabular}
\end{center}
\caption{Average coalescence time of a whole population of large size knowing that two individuals chosen at random have the same genome (left) or different genomes (right). Without conditionning on the genomes of the two individuals, the average coalescence time would be $\langle T \rangle =2$. }\label{fig:tknow2moy}
\end{figure*}

The $\theta$ dependence of this average coalescence time is shown in figure \ref{fig:tknow2moy}. Although $Y$ is a rough estimator of the genetic diversity and we consider only information about two individuals, $\langle T | \text{2 id.}\rangle$ is shifted up to 5\% compared to the case of no information.

%-----------------------------------------------------------------

One can write down a general expression for the Laplace transform $\hat{p}_{m,n}(s)=\int_0^\infty p_{m,n}(t)e^{-st}dt$ of the solution of (\ref{1group:p_rec})~:
\begin{equation}\label{psi_nm}
\hat{p}_{m,n}(s) = \sum_{1\leq n_1<\ldots<n_m=n}
\Big(\frac{B_{n,m}(n_j)}{S(n)}\Big)\times\Big( \prod_{i=2}^n
\hat{f}_i(\{n_j\},s) \Big)
\end{equation}
with functions $f_i(\{n_j\},s)$ defined as
\begin{equation} \label{def:fipjs}
\hat{f}_i(n_j,s) =
  \begin{cases}
    \frac{c_i}{s+c_i+j\theta}  & \text{for $n_j\geq i\geq n_{j-1}+1$ and $j\geq 1$}, \\
    \frac{c_i}{s+c_i}  & \text{for $n_1\geq i$}.
  \end{cases}
\end{equation}
and amplitudes~:
$$B_{n,m}(n_j) = \frac{S(m) (n+m-1)!(n-m)!}{2^{n-1}} \prod_{j=1}^{m-1}
\frac{1}{c_{n_j+j+1}}$$

This result can be obtained by counting trees and averaging on coalescence times $\tau_i$ as shown in
appendix \ref{measgenstruct}. Let us sketch briefly the derivation of (\ref{psi_nm}). The genealogy of the group of $n$ individuals can be divided into several parts which correspond to a constant number of ancestors of the subgroup of $m$ individuals, i.e. the parts are separated by coalescences \emph{among} the ancestors of the $m$ individuals. The indices $n_j$ in (\ref{psi_nm}) are the number of ancestors of the $n$ individuals at the times of these coalescences, i.e. when the number of ancestors of the $m$ individuals decreases from $j+1$ to $j$ due to a coalescence. The quantity $B_{n,m}(n_j)S(n+m)$ counts the number of trees sharing the same parameters $n_j$ and thus the sum over the $n_i$ in (\ref{psi_nm}) is an average over the shape of the trees. The value of $B_{n,m}(n_j)$ can be obtained by counting at each coalescence the number of possibilities compatible with the value $n_j$.

Given a set of parameters $n_i$, we now consider the distribution of the coalescence times $\tau_i$ conditioned by the shape of the tree and the genomes of the subgroup of $m$ individuals. Mutations are forbidden in the subtree of the $m$ individuals. Thus, if the number of ancestors of the $m$ individuals is $j$ during $\tau_i$, the probability that no mutation occur is $e^{-j\theta\tau_i}$. If one introduces the parameters $n_i$, the probability that the delay between the $(i-1)$-th coalescence and the $i$-th is $\tau_i$ and that no mutations occur on the lineages of the $m$ individuals with the same genome is $f_i(n_j,t)$ defined as~:
\begin{equation}
f_i(n_j,\tau_i) =
  \begin{cases}
    c_i e^{-(c_i+j\theta)\tau_i}  & \text{for $n_j\geq i\geq n_{j-1}+1$ and $j\geq 1$}, \\
    c_i e^{-c_i\tau_i}  & \text{for $n_1\geq i$}.
  \end{cases}
\end{equation}
The Laplace transform of these expressions gives the result (\ref{def:fipjs}) and the product of the $\hat{f}_i$ in (\ref{psi_nm}) corresponds to the average on the $\tau_i$'s. 

%---------------------------------------------------------------------

\begin{figure}[t]
\begin{center}
\input{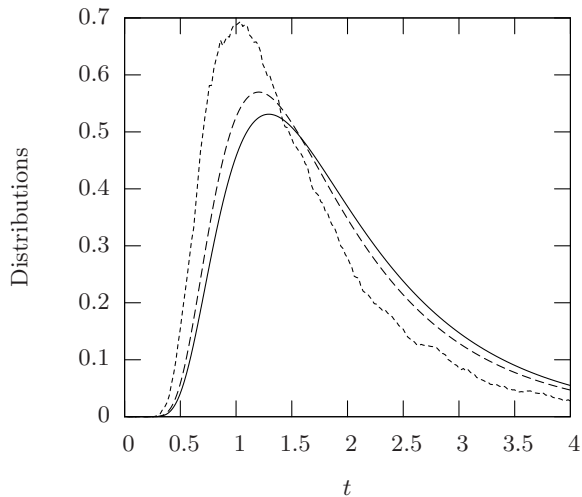}
\end{center}
\caption{Probability distribution of the coalescence time of a group of individuals knowing that the first $m$ of them have the same genome. Solid line~: stationary distribution for a large group without information. Dashed (long)~: $\theta=0.5$ and $m=2$ in a large group. Dashed (short)~: $\theta=0.5$ and $m=5$ (numerical simulations for a population of 50 individuals).
}\label{fig:tknowing2}
\end{figure}

Figure \ref{fig:tknowing2} shows the distribution $p_{2,\infty}(t)/\hat{p}_{2,\infty}(0)$ of the conditioned coalescence time $T$ obtained from (\ref{hatp2infty}). It also shows numerical results on
a population of 50 individuals which agree with analytical calculations showing how information about five
individuals modifies the coalescence time of the whole population significantly.

\section{Conclusion}

In the present paper, we have shown that the evolution of all the coalescence times at the top of the genealogical tree can be described by a Markov process (section \ref{subsec:stoch}). This Markov process allowed us to calculate various properties (\ref{T0autocorr:casc}, \ref{res:G-}, \ref{G-Gst}, \ref{DHcorr:theo}) of the age of the MRCA, in particular its autocorrelation function (\ref{T:autocorr}, \ref{T:coeffAp}). We have also shown how to calculate the correlation between the age of the MRCA and a parameter representing the genetic diversity (section \ref{corrdiv}). Our general formula (\ref{psi_nm}), correlating the age of the MRCA of $n$ individuals knowing that a sample of $p$ of them chosen at random have the same allele, is not easy to manipulate. Its interpretation as a weighted sum over a large number of tree configurations may however allow numerical simulations with Monte-Carlo methods \cite{fuli} by sampling efficiently the terms of the sum. 

The Markov property of the genealogies is the most promising result of this paper and one may hope to construct more general Markov processes of this type. A first direction would be to try to incorporate the genetic diversity in the Markov process~: whereas section \ref{corrdiv} leads only to the stationary correlation coefficients $\langle Y T^k\rangle$, the construction of a joint Markov process for the times $\tau_i$ and the sizes of the families may lead to correlations at different times and establish links between extinctions and variations of the genetic diversity. Moreover this could be related to works such as \cite{drummond} in the case where sampling the DNA of individuals at different times is possible.

Extensions of the Markov process to more realistic models would also be interesting but many aspects of the calculations may differ. For example, the shape of the genealogical trees changes in presence of selection since multiple coalescences \cite{pitman,durrett2,schweinsberg} have to be included and this should change the weights of the trees and the probabilities of extinctions of families. The study of structured populations \cite{wilkinson,wakeleyaliacar} shows that demographic and geographic effects are important~: it would be interesting to know if the Markov property of the coalescence times persists, up to changes in the transition rates. Diploidy \cite{chang,manrubia} is more problematic since it has more radical effets (\emph{e.g.} the age of the MRCA scales as $\log N$ and not as $N$ anymore) because genealogical trees have a more complicated structure with loops.

Lastly, it would be interesting to see how more detailed information about the genomes could lead to a more accurate estimation of the age of the MRCA. Analysis of section \ref{corrdiv} deals with only one gene. Distinct genes may evolve in different ways since the MRCA and, in the present generation, one is left with different parameters $Y$ for each gene. Information about the genetic diversities for different genes would modify the distribution of the times $\tau_i$ in order to account for possible differences in the number of mutations of each gene. Moreover, in real cases, the observation of different genes along a DNA sequence would be incomplete if recombination \cite{hein,wiuf,hudson} is not taken into account. Recombination acts as if the two genes of a given individuals are not inherited from the same parent. It implies that the genealogical trees of the two genes will have some different branches and the MRCA may be different for the two genes and the difference of ages between these ancestors may be worth further investigations.

\acknowledgments

D.S. thanks Professors Yamazaki Yoshihiro and Ishiwata Shin'ichi
at Waseda University (T\={o}ky\={o}, Japan) for their hospitality
during summer 2004.

\appendix

\section{Measure on genealogies}
\label{measgenstruct}

In this appendix we recall briefly the derivation of the statistics \cite{kingman,kingman2,derridapeliti} of the coalescence times in the genealogies of a set of individuals. The problem can be divided into two aspects~: the distribution of the coalescence times and the shape of the tree.

We consider a group of $n$ individuals undergoing coalescences
until they reach their MRCA. Each coalescence is characterized by
two quantities~: the waiting time until it occurs and the pair 
of individuals which coalesce.

For a large size $N$ of the population, coalescences occur one
after another. At each generation, the probability of a
coalescence between a given pair of individuals is $1/N$ and the
total probability of observing a coalescence is $c_n/N$ for a
group of $n$ individuals, where the coefficient $c_n$ is defined
as $c_n=n(n-1)/2$ (see (\ref{def:ci})). The probability of observing the first
coalescence at generation $G$ in the past is then
$\rho_n(G)=c_n/N (1-c_n/N)^G$ which becomes for the rescaled time $\tau=G/N$~:
\begin{equation}\label{waiting_coal_time} \rho_n(\tau) = c_n e^{-c_n \tau} \end{equation}

After this coalescence, we are left with $n-1$ individuals and the
rescaled time $\tau_{n-1}$ before the next coalescence is then given by
$\rho_{n-1}(\tau_{n-1})$ and so on. So, the distribution of the $(n-1)$
waiting times $\tau_i$ between two successive coalescences for a
group of $n$ individuals is~: \begin{equation}
P_n(\tau_{n},\ldots,\tau_2) = \prod_{i=2}^n
c_ie^{-c_i\tau_i}\end{equation} 

Consequently, the total
coalescence time can be written as a sum of $n-1$ independent
variables~: \begin{equation} T_n =\sum_{i=2}^n
\tau_i\end{equation}

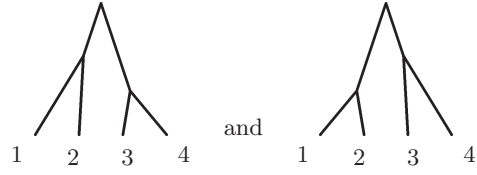
\begin{figure} 
\begin{center}
\begin{fmffile}{arbre3}
\begin{fmfgraph*}(50,50)
\fmfstraight \fmftop{i1} \fmfstraight \fmfbottom{a1,a2,b1,b2}
\fmflabel{$1$}{a1} \fmflabel{$2$}{a2} \fmflabel{$3$}{b1} \fmflabel{$4$}{b2}
\fmf{plain}{i1,a}\fmf{plain}{i1,b} \fmf{plain}{a,aa1}
\fmf{plain}{a,aa2} \fmf{plain}{aa1,aaa1} \fmf{plain}{aa2,aaa2}
\fmf{plain}{aaa1,a1} \fmf{plain}{aaa2,a2}\fmf{plain}{b,b1}
\fmf{plain}{b,b2}
\end{fmfgraph*} \quad\quad and \quad\quad
\begin{fmfgraph*}(50,50)
\fmfstraight \fmftop{i1} \fmfstraight \fmfbottom{a1,a2,b1,b2}
\fmflabel{$1$}{a1} \fmflabel{$2$}{a2} \fmflabel{$3$}{b1} \fmflabel{$4$}{b2}
\fmf{plain}{i1,a}\fmf{plain}{i1,b} \fmf{plain}{a,a1}
\fmf{plain}{a,a2} \fmf{plain}{b,bb1} \fmf{plain}{b,bb2}
\fmf{plain}{bb1,bbb1} \fmf{plain}{bb2,bbb2} \fmf{plain}{bbb1,b1}
\fmf{plain}{bbb2,b2}
\end{fmfgraph*}
\end{fmffile}
\end{center}
\caption{\label{fig:treestruct}Two genealogical trees. Their topologies are identical but the chronology is different.}
\end{figure}

Once the dates of the coalescences are known, we have to decide which branches coalesce at each step. We will consider here that a tree $\mathcal{T}$ is completely characterized by its topology \emph{and} the chronological order of these $n-1$ coalescences. With this definition which is convenient for our calculations, the two trees shown in figure \ref{fig:treestruct} are distinct.

The total number of such ordered trees is thus
\begin{equation} S(n) = \prod_{i=2}^n c_i = \frac{n!(n-1)!}{2^{n-1}}
\end{equation} and they are all equally likely. The probability measure $\mu_n$
of a given genealogy factorizes as~: 
\begin{equation}\label{mu} \mu_n(\mathcal{T},\{\tau_i\}) = \frac{1}{S(n)} \rho_n(\tau_n)\rho_{n-1}(\tau_{n-1})\ldots \rho_2(\tau_2) \end{equation}

\bigskip

For a given tree, one can determine from (\ref{mu}) for each ancestor on a branch of the tree, the distribution of its number of descendants in the present generation. For example, right before the last coalescence, the ancestors of the group of size $n$ consists of two parents who have in the present generation $p$ and $n-p$ descendants respectively. The sizes $p$ and $n-p$ of these two groups can be obtained by
counting the number $s(n,p)$ of trees satisfying this constraint.
The probability $\rho_n(p)$ of observing the subdivision $(p,n-p)$
with $1\leq p\leq n-1$ is given by~:
\begin{equation}\label{def:rhon}
\rho_n(p) = \frac{s(n,p)}{S(n)} =
\frac{1}{S(n)}S(p)S(n-p)\binom{n}{p}\binom{n-2}{p-1}
\end{equation}
The binomial coefficient $\binom{n}{p}$ counts the number of ways
of making the groups of $p$ and $n-p$ individuals, the 
coefficients $S(p)$ and $S(n-p)$ count the number of subtrees for
each groups and the factor $\binom{n-2}{p-1}$ counts the ways of
organizing the chronological order between the coalescences of the two subtrees. The
dependence on $p$ disappears in (\ref{def:rhon}) and
$\rho_n(p)$ is the uniform distribution~:
\begin{equation}\label{res:rhon}
\rho_n(p) = \frac{1}{n-1}
\end{equation}

One should notice that this result is obtained for a large population $N$ and a group of size $n\ll N$, such that coalescences occur only between pairs of individuals and not more. However, if $n$ is large enough and if we define the density $x=p/n$, the corresponding distribution $\rho(x)$ is uniform on $[0,1]$.

\bigskip

For a branch of
length $\tau$, the number $m$ of mutations has a Poisson
distribution : \[ P(\tau,m) =
\frac{(\theta\tau)^m}{m!}e^{-\theta\tau} \] So the probability of
observing no mutation on this branch, which is the only relevant quantity in the infinite allele case) is given by :
\begin{equation} P_\text{no mut}(\tau)= e^{-\theta\tau} \end{equation}

\section{Dynamics of the times $\tau_i$}
\label{app:cascade}

Figure \ref{fig:evoltau2} shows the stochastic dynamics of the coalescence time $\tau_2$. Actually, all the elementary coalescence times $\tau_i$ of figure \ref{fig:arbre} defined in appendix \ref{measgenstruct} have similar dynamics~: either they increase by $\tau_{i+1}$ or they are reset to $\tau_{i+1}$. The idea of a Markov process in genealogies is not new and some features are presented in \cite{donnelly}.

If one considers a generic tree as shown in figure \ref{fig:arbre} truncated below $\tau_n$, one sees that the times $\tau_i$ topple when some lineages coming from the $n$ ancestors at the "leaves" of the truncated tree disappear. Let us assume that the lineage of a given ancestor among these $n$ disappears and that this ancestor is directly connected to the $j$-th coalescence, i.e. the coalescence separating $\tau_j$ and $\tau_{j+1}$. For example, if $j=1$, the ancestor is directly connected to the MRCA and if $j=2$, it is directly connected to $A_1$ in figure \ref{fig:arbre}. If the lineage of this ancestor in the present generation disappears, the times $\tau_i$ topple at rank $j+1$, i.e. they are redefined as~:
$$ \begin{cases}
\tau'_i = \tau_i & \text{for $i<j$}\\
\tau'_j = \tau_j+\tau_{j+1} & \\
\tau'_i = \tau_{i+1} & \text{for $i>j$}
\end{cases} $$

Let us call $P_\text{node}(n,j)$ the probability that a given ancestor among the $n$ is directly connected to the node of the $j^\text{th}$ coalescence, i.e. the lineage of this ancestor does not participate at any coalescence until the number of ancestor reaches $j$. With these notations, the probability $p_j dt$ defined in (\ref{prob:cascade}) that the times topple at rank $j$ is the probability that the lineage which disappears during $dt$ (probability $\alpha_n dt$) is the lineage connected to the $j^\text{th}$ coalescence and thus it is given by~:
\begin{equation}\label{pi:decomp}
p_j dt = P_\text{node}(n,j) \alpha_n dt 
\end{equation}

The value of $\alpha_n$ can be derived by introducing the probability $Q_t(n,t_0)$ that the number of ancestors at time $t_0<t$ of the whole population at time $t$ is $n$. These $n$ ancestors at time $t_0$ generate all the population at time $t$ which can be divided in $n$ groups, each depending on the ancestor they come from. At time $t+dt$, either one of these groups gets extinct (probability $\alpha_n dt$) and the number of ancestors at $t_0$ of the population at $t+dt$ is $n-1$, or this number is still equal to $n$. The probabilities $Q_t(n,t_0)$ satisfy the following equations~:
\begin{equation*}
Q_{t+dt}(n-1, t_0) = Q_{t}(n,t_0) \alpha_n dt + Q_t(n-1,t_0)  (1-\alpha_{n-1} dt)
\end{equation*}
It gives the differential equation~:
\begin{equation}\label{Plignee_ext}
\frac{d}{dt}Q_t(n,t_0) = Q_t(n,t_0) \alpha_n - Q_t(n-1,t_0) \alpha_{n-1}
\end{equation}
In the stationary regime, this probability is $Q_t(m,t-\tau)=z_n(\tau)$ where the $z_m$'s have been defined in section \ref{subsec:stoch} and satisfy (\ref{zm:rec}). Comparing (\ref{Plignee_ext}) and (\ref{zm:rec}) leads to~:
\begin{equation}\label{alphan:res}
\alpha_n = c_n
\end{equation}

The probability $P_\text{node}(n,j)$ is the probability that, in the genealogy of a group of $n$ individuals, the lineage of a given individual among the $n$ is directly connected to the node of the $j$-th coalescence, i.e. that coalescences do not involve its lineage until the number of ancestors of the group is reduced to $i+1$. Counting the number of possibilities for each coalescence gives~:
\begin{equation}\label{Pnode:res}
P_\text{node}(n,j) = \frac{j}{c_{j+1}}\prod_{k=j+2}^n\frac{c_k-(k-1)}{c_k} = \frac{j}{c_n}
\end{equation}

Putting (\ref{Pnode:res}) and (\ref{alphan:res}) in (\ref{pi:decomp}) gives the toppling rates presented in  (\ref{prob:cascade})~:
\begin{equation}
p_i = \frac{i}{c_n} \times c_n = i
\end{equation}

\bibliographystyle{unsrt}
\bibliography{bibliogenea}

\end{document}